\documentclass[10pt]{iopart}
\makeatletter
\newenvironment{figurehere}
{\def\@captype{figure}} {} \makeatother
\usepackage{graphicx}
\usepackage{graphicx}
\usepackage{cite}
\begin{document}
\title{Thermal Entanglement of a Spin-1/2 Ising-Heisenberg Model on a Symmetrical Diamond Chain}
\author{N. S. Ananikian$^{1, 4}$, L. N. Ananikyan$^1$, L. A. Chakhmakhchyan$^{1,2,3}$, and Onofre Rojas$^{4}$\\[1mm]
{\small \sl $^1$A.I. Alikhanyan National Science Laboratory, 0036 Yerevan, Armenia,} \\
{\small \sl $^2$ Institute for Physical Research, 0203 Ashtarak-2, Armenia}\\
{\small \sl $^3$ Laboratoire Interdisciplinaire Carnot de Bourgogne, UMR 5209 CNRS - Universit\'{e} de Bourgogne, Dijon, France}\\
{\small \sl $^4$ Departamento de Ciencias Exatas, Universidade Federal de Lavras, CP 3037, 37200000, Lavras, MG, Brazil}}
\begin{abstract}
The entanglement quantum properties of a spin-$1/2$ Ising-Heisenberg
model on a symmetrical diamond chain were analyzed. Due to the
separable nature of the Ising-type exchange interactions between
neighboring Heisenberg dimers, calculation of the entanglement can be
performed exactly for each individual dimer. Pairwise thermal entanglement was studied in terms of the isotropic Ising-Heisenberg model, and analytical expressions for the concurrence (as a measure of bipartite entanglement) were obtained. The effects
of external magnetic field $H$ and next-nearest neighbor interaction $J_m$
between nodal Ising sites were considered. The ground-state structure
and entanglement properties of the system were studied in a
wide range of the coupling constant values.
Various regimes with different values of the ground-state entanglement
were revealed, depending on the relation between competing interaction strengths. Finally, some novel effects, such as the two-peak
behavior of concurrence versus temperature and coexistence of phases
with different values of magnetic entanglement were observed.
\begin{description}
\item[PACS numbers] 75.10.Jm, 75.50.Ee, 03.67.Mn, 64.70.Tg
\end{description}
\end{abstract}

\maketitle

\section{Introduction}\label{intr}

During the last two decades low-dimensional magnetic materials with
competing interactions or geometrical frustration have become an
intriguing research object. Particularly, these materials exhibit a rich variety of unusual ground
states and thermal properties, as a result of zero and finite
temperature phase transitions \cite{frust, compete, diep, thermal,
thermal1}. As attractive models among these systems, one should mention the ones, having a
diamond-chain structure. The latter consists of diamond-shaped topological
units along the chain (Fig.~\ref{chain}). It has been observed that
the compounds A$_3$Cu$_3$(PO$_4$)$_4$ with A=Ca, Sr \cite{diam1},
Bi$_4$Cu$_3$V$_2$O$_{14}$ \cite{diam2} and Cu$_3$(TeO$_3$)$_2$Br$_2$
\cite{diam3} can be nicely modeled by the Heisenberg diamond chain.
Besides, recent
experimental results on the natural mineral azurite
(Cu$_3$(CO$_3$)$_2$(OH)$_2$) \cite{azurite} showed that Cu$^{2+}$
ions of this material form a spin-$1/2$ diamond chain. Furthermore,
the discovery of a plateau at $1/3$ of the saturation value in the
low-temperature magnetization curve \cite{azurite, azurite1} has
triggered an intensive interest in the magnetic properties of
azurite \cite{azexp, aztheor}. Azurite falls into the class of
geometrically frustrated magnets. However, the question of the strength and the type of exchange interactions for this natural mineral, despite the long-standing interest, is still open. The first diamond spin chain was
explored under a symmetrical condition $J_1=J_3$ \cite{first} that
predicted magnetization plateaus both at $1/3$ and $1/6$ of
saturation \cite{predict}. The frustrated diamond chain
with ferromagnetic interactions $J_1, J_3<0$ and antiferromagnetic
interaction $J_2>0$ was also investigated theoretically
\cite{another}.  
Other exchange interactions, like an additional cyclic four-spin \cite{cyclic} and $J_m$ interaction between monomeric units
(the so called generalized diamond chain) \cite {new1} were considered. Additionally, the importance of an anisotropic exchange and Dzyaloshinskii-Moriya interaction \cite{new} or interchain coupling
\cite{inter} was discussed. To sum up, note that the theory predictions
for certain values of exchange coupling constants within a
relatively broad range can fit the experimental results. The
controversy on these values seems to be cleared up only recently (the
latest comparison of experimental and theoretical results can be
found in Ref. \cite{richter}).

Motivated by the controversies discussed above and the fact that
different compounds can be described by means of a diamond chain, 
we shall explore
systematically the generalized symmetrical spin-1/2 diamond chain
with various competing interactions in a magnetic field. Unfortunately, the rigorous theoretical treatment of geometrically
frustrated quantum Heisenberg models is difficult to fulfil. The problem arises due to a
non-commutability of spin operators involved in the Heisenberg
Hamiltonian. This is also a primary cause of a presence of quantum
fluctuations. Owing to this fact, we will use the recently proposed
geometrically frustrated Ising-Heisenberg diamond chain model
\cite{symm, symm1, lusn}. The latter suggests to overcome the mathematical
difficulties by introducing the Ising spins at the nodal sites and
the Heisenberg dimers on the interstitial decorating sites of the
diamond chain (Fig.~\ref{chain}). For understanding of the
properties of underlying purely quantum models it is required to
obtain an analytic expression for all thermodynamic functions of the
model. Note that some exactly solvable models with Ising and
Heisenberg bonds can also provide satisfactory quantitative picture
\cite{quant}.

In the present paper we shall mainly deal with the quantum
entanglement properties of the spin-1/2 Ising-Heisenberg model on a
generalized symmetrical diamond chain. It is well-known, that the
entanglement is a generic feature of quantum correlations in
systems, that cannot be quantified classically \cite{review,
review1}. It provides a new perspective for understanding the
quantum phase transitions (QPTs) and collective phenomena in
many-body and condensed matter physics. This problem, which has been
under scrutiny for nearly two decades, has recently attracted much
attention~\cite{entangle, entangle1, entangle2, entangle3,
entangle4}. A new line research points to a relation between the
entanglement of a many-particle system and the existence of the QPTs
and scaling \cite{QPT1, alba}. On the other hand, the study of
entanglement in solid state physics is of a great relevance to the
area of Quantum Information and Quantum Computation, since many
proposals of quantum chips are solid state-based. Although it was
believed that the entanglement should not manifest itself in
macroscopic objects (because of a large number of constituents
interacting with the surroundings that induce the decoherence
phenomena
), it was theoretically
demonstrated that entangled states can exist in solids at finite
temperatures. This kind of entanglement is referred to in
literature as "the thermal entanglement" \cite{entangle}. And
afterwards a few experimental evidences have been reported for
low-dimensional spin systems \cite{exp}, confirming the presence of
entanglement in solid state materials.

Returning to the spin-1/2 Ising-Heisenberg model on a diamond
chain we remark that the nodal Ising spins represent a barrier
for quantum fluctuations.
On the other hand, taking into account that each
Heisenberg dimer interacts with its neighboring dimer through the
Ising-type, i.e. classical  exchange interaction, we find that the
states of two adjacent dimers become separable (disentangled)
\cite{review}. Thus, we can calculate the concurrence (the measure
of pairwise entanglement \cite{wooters}), which characterizes
quantum features of the system, for each dimer separately. The main
objective of the paper is to reveal different regimes of the
symmetrical diamond chain and to analyze new quantum effects (such
as double peak behavior in the concurrence versus temperature
curves, existence of magnetic entanglement \cite{entangle} of two
different values). 

The rest of the paper is organized as follows: we start in Sec.~\ref{method} by obtaining concurrence as a measure of entanglement of the spin-$1/2$ Ising-Heisenberg model on a generalized symmetrical diamond chain. The ground-state structure and the entanglement features of the ideal diamond chain ($J_1=J_3$, $J_m=0$) are discussed in Sec.~\ref{zero}. The following section contains similar
results with the incorporation of $J_m$ interaction. Some comments
and concluding remarks are given in Sec.~\ref{concl}.

\section{Concurrence and thermal entanglement of the spin-$1/2$ Ising-Heisenberg model on a generalized symmetrical diamond}\label{method}
\begin{figure}
\begin{center}
\includegraphics[width=8cm]{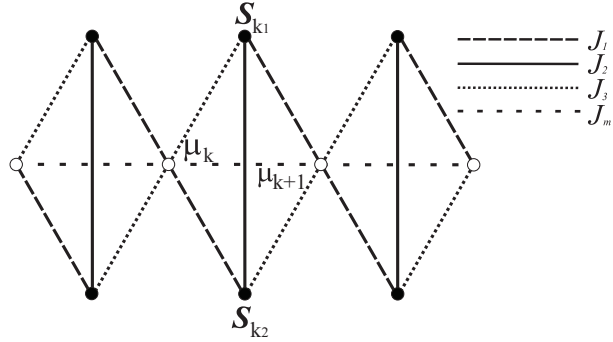}
\caption{\small{ A cross-section of a generalized symmetrical diamond chain ($k$ labels the number of the cluster).
The empty (monomeric units) and full circles (dimeric units) denote lattice positions of the Heisenberg and Ising spins (within the proposed Ising-Heisenberg model), respectively. Solid lines schematically reproduce the Heisenberg $J_{2}$ interactions between dimeric units, while the broken ones label the Ising-type (nearest-neighbor $J_1, J_3$ and next-nearest neighbor $J_m$) interactions.
\label{chain}}}
\end{center}
\end{figure}

We consider the spin-$\frac{1}{2}$ Ising-Heisenberg model on
a generalized
symmetrical diamond chain ($J_1=J_3=J$), which
consists of monomeric and dimeric sites (empty and full circles in
Fig.~\ref{chain}, respectively). Within the proposed Ising-Heisenberg
model, the monomeric (nodal) sites are occupied by Ising spins,
while the dimeric sites by Heisenberg-type spins. The Hamiltonian
can be written as follows:

\begin{eqnarray}
\mathcal{H}=\sum_{k=1}^{N}\mathcal{H}_k=&&\sum_{k=1}^{N}\left[J_{2}\mathbf{S}_{k_1}\mathbf{S}_{k_2}+
J(\mu^z_{k}+\mu^z_{k+1})(S^z_{k_1}+S^z_{k_2})+\right. \nonumber\\
&&+J_m \mu^z_{k}\mu^z_{k+1}-H\left.\left(S^z_{k_1}+S^z_{k_2}+\frac{\mu^z_{k}+\mu^z_{k+1}}{2}\right)\right], \label{1}
\end{eqnarray}
where the summations run over clusters (Fig.~\ref{chain}),
$\mathcal{H}_k $ represents the Hamiltonian of the $k-th$ cluster,
$\mathbf{S}_k=(S^x_k, S^y_k, S^z_k)$ denotes the Heisenberg
spin-$\frac{1}{2}$ operator, $\mu_k$ is the Ising spin. Considering,
that each Ising spin belongs simultaneously to two clusters, we have
taken a 1/2 factor for the Ising spins in the last term of (1), which
incorporates the effects of external magnetic field. $J, J_2, J_m>0$ corresponds to the antiferromagnetic couplings. The system
will be strongly frustrated due to the chain's geometry and
existence of competing interactions $J, J_2$  and $J_m$. When $J_m=0$ we deal with the so called ideal diamond chain
\cite{symm}.
Before introducing the calculations and discussion
we would like to emphasize the fact which was already discussed in
Sec.~\ref{intr}: the states of two neighboring Heisenberg dimers
(with interaction $J_2$) are separable (disentangled), because of a
classical character of the coupling between them (by means of the
Ising spin). Hence we can calculate the entanglement for each of the
dimers individually.
Note, that a different approach of a mean-field-like treatment, based on the Gibbs-Bogoliubov inequality was used in Ref.~\cite{mik},
where all the couplings between the diamond chain sites were chosen to be of a quantum (Heisenberg-type) character.

Here we use concurrence (as a measure of bipartite entanglement \cite{wooters}) of the Heisenberg
dimers by tracing out the Ising spins in each cluster.
For the construction of eigenvectors
of each cluster we will take into account that $\mathcal{H}_k$
possesses a symmetry corresponding to the permutations
$\mu_{k}\leftrightarrow \mu_{k+1}$ and $\{\mu_{k}\leftrightarrow
\mu_{k+1}; \mathbf{S}_{k_1}\leftrightarrow\mathbf{S}_{k_2}\}$.
Besides, the Hilbert space of the cluster $\mathcal{H}_{cluster}$
can be presented as
$\mathcal{H}_{cluster}=\mathcal{H}_{k}\otimes\mathcal{H}_{dimer}\otimes\mathcal{H}_{k+1}$,
where $\mathcal{H}_{k_1}$, $\mathcal{H}_{dimer}$,
$\mathcal{H}_{k_2}$ denotes the Hilbert spaces of $\mu_{k}$,
Heisneberg dimer and $\mu_{k+1}$ respectively. We obtain the
following eigenvectors, due to the symmetries and Hilbert space structure
(hereafter, the letter $k$ labels the number of the cluster):
\begin{eqnarray}\nonumber
&|\psi_1\rangle=\frac{1}{\sqrt{2}}(|\uparrow_{k_1}\otimes\downarrow_{k_2}\rangle+|\downarrow_{k_1}\otimes\uparrow_{k_2}\rangle)\otimes
|\uparrow_{k}\uparrow_{k+1}\rangle;\\ \nonumber
&|\psi_2\rangle=\frac{1}{\sqrt{2}}(|\uparrow_{k_1}\otimes\downarrow_{k_2}\rangle+|\downarrow_{k_1}\otimes\uparrow_{k_2}\rangle)
\otimes(|\uparrow_{k}\downarrow_{k+1}\rangle+|\downarrow_{k}\uparrow_{k+1}\rangle); \\ \nonumber
&|\psi_3\rangle=\frac{1}{\sqrt{2}}(|\uparrow_{k_1}\otimes\downarrow_{k_2}\rangle+|\downarrow_{k_1}\otimes\uparrow_{k_2}\rangle)
\otimes(|\uparrow_{k}\downarrow_{k+1}\rangle-|\downarrow_{k}\uparrow_{k+1}\rangle); \\ \nonumber
&|\psi_4\rangle=\frac{1}{\sqrt{2}}(|\uparrow_{k_1}\otimes\downarrow_{k_2}\rangle+|\downarrow_{k_1}\otimes\uparrow_{k_2}\rangle)\otimes
|\downarrow_{k}\downarrow_{k+1}\rangle;\\ \nonumber
&|\psi_5\rangle=\frac{1}{\sqrt{2}}(|\uparrow_{k_1}\otimes\downarrow_{k_2}\rangle-|\downarrow_{k_1}\otimes\uparrow_{k_2}\rangle)\otimes
|\uparrow_{k}\uparrow_{k+1}\rangle;\\ \nonumber
&|\psi_6\rangle=\frac{1}{\sqrt{2}}(|\uparrow_{k_1}\otimes\downarrow_{k_2}\rangle-|\downarrow_{k_1}\otimes\uparrow_{k_2}\rangle)
\otimes(|\uparrow_{k}\downarrow_{k+1}\rangle+|\downarrow_{k}\uparrow_{k+1}\rangle); \\ \nonumber
&|\psi_7\rangle=\frac{1}{\sqrt{2}}(|\uparrow_{k_1}\otimes\downarrow_{k_2}\rangle-|\downarrow_{k_1}\otimes\uparrow_{k_2}\rangle)
\otimes(|\uparrow_{k}\downarrow_{k+1}\rangle-|\downarrow_{k}\uparrow_{k+1}\rangle); \\ \nonumber
&|\psi_8\rangle=\frac{1}{\sqrt{2}}(|\uparrow_{k_1}\otimes\downarrow_{k_2}\rangle-|\downarrow_{k_1}\otimes\uparrow_{k_2}\rangle)\otimes
|\downarrow_{k}\downarrow_{k+1}\rangle\\ \nonumber
&|\psi_9\rangle=|\uparrow_{k_1}\otimes\uparrow_{k_2} \rangle\otimes|\uparrow_{k}\uparrow_{k+1}\rangle;\\
&|\psi_{10}\rangle=\frac{1}{\sqrt{2}}
|\uparrow_{k_1}\otimes\uparrow_{k_2}\rangle\otimes(|\uparrow_{k}\downarrow_{k+1}\rangle+|\downarrow_{k}\uparrow_{k+1}\rangle);\\ \nonumber
&|\psi_{11}\rangle=\frac{1}{\sqrt{2}}
|\uparrow_{k_1}\otimes\uparrow_{k_2}\rangle\otimes(|\uparrow_{k}\downarrow_{k+1}\rangle-|\downarrow_{k}\uparrow_{k+1}\rangle);\\ \nonumber
&|\psi_{12}\rangle=|\uparrow_{k_1}\otimes\uparrow_{k_2} \rangle\otimes|\downarrow_{k}\downarrow_{k+1}\rangle;\\ \nonumber
&|\psi_{k_{13}}\rangle=|\downarrow_{k_1}\otimes\downarrow_{k_2} \rangle\otimes|\uparrow_{k}\uparrow_{k+1}\rangle;\\ \nonumber
&|\psi_{14}\rangle=\frac{1}{\sqrt{2}}
|\downarrow_{k_1}\otimes\downarrow_{k_2}\rangle\otimes(|\uparrow_{k}\downarrow_{k+1}\rangle+|\downarrow_{k}\uparrow_{k+1}\rangle);\\ \nonumber
&|\psi_{15}\rangle=\frac{1}{\sqrt{2}}
|\downarrow_{k_1}\otimes\downarrow_{k_2}\rangle\otimes(|\uparrow_{k}\downarrow_{k+1}\rangle-|\downarrow_{k}\uparrow_{k+1}\rangle);\\ \nonumber
&|\psi_{16}\rangle=|\downarrow_{k_1}\otimes\downarrow_{k_2} \rangle\otimes|\downarrow_{k}\downarrow_{k+1}\rangle;\\  \nonumber\label{2}
\end{eqnarray}
and the corresponding eigenvalues:
\begin{eqnarray} \nonumber
&E_1=\frac{1}{4} \left(-2 H+J_m+J_2\right); \quad E_2=E_3=-\frac{J_m-J_2}{4}; \\ \nonumber
&E_4= \frac{1}{4} (J_m+J_2+2 H); \quad E_5=\frac{1}{4} (-2 H+J_m-3 J_2);  \\ \nonumber
&E_6=E_7= -\frac{J_m+3 J_2}{4}; \quad E_8=\frac{1}{4} (2 H+J_m-3 J_2);\\
&E_9=-\frac{3 H}{2}+\frac{J_m+J_2}{4}+J; \quad E_{10}=E_{11}=-H-\frac{J_m-J_2}{4}; \\ \nonumber
&E_{12}= -\frac{H}{2}+\frac{1}{4} \left(J_m+J_2-4 J\right); \quad E_{13}=\frac{H}{2}+\frac{1}{4} \left(J_m+J_2-4 J\right); \\ \nonumber
&E_{14}= E_{15}=\frac{1}{4} (-J_m+J_2+4 H); \quad E_{16}=\frac{3 H}{2}+\frac{J_m+J_2}{4}+J. \\  \nonumber\label{3}
\end{eqnarray}

We study \textit{concurrence} $C(\rho)$, to quantify pairwise
entanglement \cite{wooters}, defined as
\begin{equation}
C(\rho)=max\{\lambda_1-\lambda_2-\lambda_3-\lambda_4, 0\}, \label{4}
\end{equation}
where $\lambda_i$'s are the square roots of the eigenvalues of the
corresponding operator for the density matrix
\begin{equation}
\tilde{\rho}=\rho_{12}(\sigma_1^y\otimes\sigma_2^y)\rho_{12}^*(\sigma_1^y\otimes\sigma_2^y)
\label{5}
\end{equation}
in descending order. Since we consider pairwise entanglement, we
should use the reduced density matrix $\rho_{12}$, by tracing out
two (of four) spins of the cluster.
The reduced density matrix $\rho_{12}$ is defined as \cite{reduce}
\begin{equation}
\rho_{12}=\sum_{\alpha}\langle\alpha|\rho|\alpha\rangle. \label{5.2}
\end{equation}
In this equation $|\alpha\rangle$ denotes basis vectors of the Hilbert space associated with the system,
with respect to which the density matrix is reduced. The summation runs over all these basis vectors.
Since in our case we make reduction with respect to two spins, $|\alpha\rangle=\{|\downarrow\downarrow\rangle, |\downarrow\uparrow\rangle,
 |\uparrow\downarrow\rangle, |\uparrow\uparrow\rangle\}$.

It is obvious that the only
entangled pair is formed by the Heisenberg spins. Other pairs are
disentangled (separable) because of the classical (diagonal)
character of the Ising-type interaction between them. Hence we will
be interested in the reduced density matrix, constructed by tracing
out two Ising-type spins $\mu_{k}$ and $\mu_{k+1}$. In other words,
$\rho_{k_{12}}=\mathrm{Tr}_{\{\mu_{k}, \mu_{k+1}\}}\rho_k$ and the full
density matrix $\rho_k$ of the $k-th$ cluster is defined as (here and further Boltzmann
constant is set to be $k_B=1$)

\begin{equation}
\rho_k=\frac{1}{Z_k}\sum_{i=1}^{16}\exp(-E_i/T)|\psi_i\rangle
\langle\psi_i|, \label{6}
\end{equation}
where $Z_k$ is the partition function:
\begin{eqnarray}\nonumber
Z_k=\mathrm{Tr}\rho_k=&e^{-\frac{6 H+J_m+4 J+J_2}{4 T}} \left(2 e^{\frac{H+J_m+2 J}{2
T}}+2
   e^{\frac{3 H+J_m+2 J}{2 T}}+\right. \\  & 2 e^{\frac{5 H+J_m+2 J}{2 T}}+2
   e^{\frac{3 H+J_m+2 J+2 J_2}{2 T}}+e^{\frac{H+J}{T}}+e^{\frac{2
   (H+J)}{T}}+ \\ \nonumber & \left.e^{\frac{2 H+J}{T}}+e^{\frac{H+2
   J}{T}}+e^{\frac{H+J+J_2}{T}}+e^{\frac{2 H+J+J_2}{T}}+e^{\frac{3
   H}{T}}+1\right). \nonumber
\end{eqnarray}
Using the definition (\ref{5.2}), the basis vectors $|\alpha\rangle=\{|\downarrow\downarrow\rangle, |\downarrow\uparrow\rangle,
 |\uparrow\downarrow\rangle, |\uparrow\uparrow\rangle\}$ we construct the reduced density matrix $\rho_{k_{12}}$ of the $k-th$ cluster:
\begin{equation}
\rho_{k_{12}}=\left(
\begin{array}{llll}
 u & 0 & 0 & 0 \\
 0 & w & y & 0 \\
 0 & y^* & w & 0 \\
 0 & 0 & 0 & v
\end{array}
\right),\label {7}
\end{equation}
where

\begin{eqnarray}
u=&2 e^{\frac{4 H+J_m-J_2}{4 T}}+e^{-\frac{-2 H+J_m-4 J+J_2}{4
   T}}+e^{-\frac{-6 H+J_m+4 J+J_2}{4 T}},\nonumber \\
v=&e^{-\frac{6 H+J_m+4 J+J_2}{4 T}} \left(2 e^{\frac{H+J_m+2 J}{2
   T}}+e^{\frac{H+2 J}{T}}+1\right),\nonumber\\
w=&\frac{1}{2} \left(e^{\frac{J_2}{T}}+1\right) e^{-\frac{2
H+J_m+J_2}{4
   T}} \left(2 e^{\frac{H+J_m}{2 T}}+e^{H/T}+1\right),\\
y=&-\frac{1}{2} \left(e^{\frac{J_2}{T}}-1\right) e^{-\frac{2
H+J_m+J_2}{4
   T}} \left(2 e^{\frac{H+J_m}{2 T}}+ e^{H/T} +1\right). \nonumber\label{8}
\end{eqnarray}
The density matrix $\rho_{k_{12}}$ in Eq.~(\ref{7}) has a form of a so called $X$-state \cite{xstate}, since the Hamiltonian $\mathcal{H}_k$ is translationary invariant with a
symmetry $[S_z, \mathcal{H}_k]=0$ ($S_z=1/2(\mu^z_{k}+\mu^z_{k+1})+S^z_{k_1}+S^z_{k_2}$) \cite{matrix}.
The concurrence $C(\rho)$ of such an $X$-state
density matrix has the following form \cite{form}:
\begin{equation}
C(\rho)=\frac{2}{Z}max(|y|-\sqrt{u v}, 0). \label{9}
\end{equation}

We note here that the reduced density matrix $\rho_{k_{12}}$ of any pair of spins, different from the Heisenberg dimer has no non-diagonal elements, responsible for the quantum correlations, i.e. entanglement (see (\ref{9})). Thus we conclude, that there is no entanglement between a pair of spins which
contains at least one Ising spin.

In Eq.~(\ref{2}), one finds a set of states with maximum value of
entanglement, for which the Heisenberg dimer is in a singlet or
a triplet state ($\psi_i$'s with $i=1,...,8$). As for the rest
of the states ($\psi_i$'s with $i=9,...,16$) the Heisenberg
dimer is in a separable state and therefore these
$\psi_i$'s are non-entangled ones.

\subsection{Ideal diamond chain}\label{zero}

\begin{figure}
\begin{center}
\includegraphics[width=8cm]{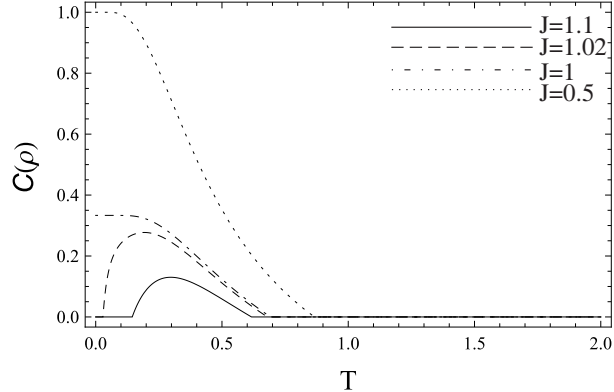}
\caption { \small {Concurrence $C(\rho)$ versus temperature $T$ for
$J_2=1$, $J_m=0$, $H=0$, and different values of $J$.}}
\label{C_T_0}
\end{center}
\end{figure}

In this section we proceed to the investigation of entanglement features of a dimeric unit of an ideal diamond chain ($J_m=0$).
First, we study the behavior of $C(\rho)$ at $H=0$. We will discuss
here three regimes, depending on the value of $J-J_2$: $J-J_2>0$,
$J-J_2<0$ and $J-J_2=0$. In the first case, as one finds out from
(3), the ground state contains two-fold degenerate states
$\psi_{12}$ and $\psi_{13}$. Since these states are
factorable, the corresponding dependency curve of $C(\rho)$ from
temperature $T$ starts from $C(\rho)=0$ (Fig.~\ref{C_T_0}).
Furthermore, the entanglement can be invoked by increasing the
temperature (for values of $J-J_2$ close to $0$). This happens since
the contribution of entangled states in the mixture $\rho_k$ increases
with the growth of temperature $T$. The local maximum, appearing here
arises due to the optimal thermal mixing of all eigenstates in the
system. This maximum becomes narrower and smaller and gradually
vanishes by increasing $J-J_2$. But the value of $J-J_2$
corresponding to disappearing of $C(\rho)$ also depends  on the
value of $J_2$ (e.g. for $J_2=1$, $J-J_2\approx0.2$). The latter
becomes obvious, if one takes into account that $J_2$, being the
coupling constant of the Heisenberg type interaction between dimeric
units, is responsible for the strength of quantum correlations
between Heisenberg spins. We would like to emphasize here that in
the case $J-J_2>0$ the system exhibits weak ($0<J_2<J$) or no
frustration ($J_2<0$).

In the second case, when $J-J_2<0$, the system will obviously
manifest more of its quantum nature. Firstly, the dependency curve
of $C(\rho)$ from temperature starts from $C(\rho)=1$ at $T=0$
(Fig.~\ref{C_T_0}). This happens due to the fact that at
zero temperature the maximum entangled states $\psi_5$,
$\psi_6$, $\psi_7$ and $\psi_8$ form
four-fold degenerate ground state with the value of $C(\rho)=1$ for
the corresponding reduced density matrix $\rho_{k_{12}}$. When the
temperature is increased, the concurrence gradually disappears
because of the thermal mixing with other states of the system
(including the factorable ones). The sudden-death temperature $T_d$,
corresponding to the dying out of quantum correlations in the system
can be found through the equation $C(\rho)=0$. It has the following
form:

\begin{equation}
x^{-J} \left(x^J+1\right)^2=2 \left|x^{J_2}-1\right|, \label{10}
\end{equation}
where $x=e^{1/T}$. The solution can be presented in the form
$T_d=J/\log a $ (when $J-J_2<0$), where $a$ depends on the ratio
parameter $J_2/J$. Increasing this ratio, $a$ decreases, but the
linear dependence on $J$ remains (e.g. when $J_2/J=2$,
$a=\frac{1}{4} (3+\sqrt{17})$).

Finally, the case $J-J_2=0$ can be regarded as a boundary case in the
following sense. Here the ground state is six-fold degenerate, containing additionally
$\psi_{12}$ and $\psi_{13}$, besides $\psi_5$, $\psi_6$,
$\psi_{7}$ and $\psi_8$ (in other words all the states as in previous two
cases). Since the $\psi_{12}$ and $\psi_{13}$ are factorable, this leads to
lower entanglement of the ground state's reduced matrix, that is $C(\rho)=1/3$ (Fig.~\ref{C_T_0}).
Moreover, the above discussed sudden-death temperature $T_d$ is lower, than in the case $J-J_2<0$ (although again $T_d=J/\log a $ with $a=2+\sqrt5$).

\begin{figure}
\begin{center}
\includegraphics[width=8cm]{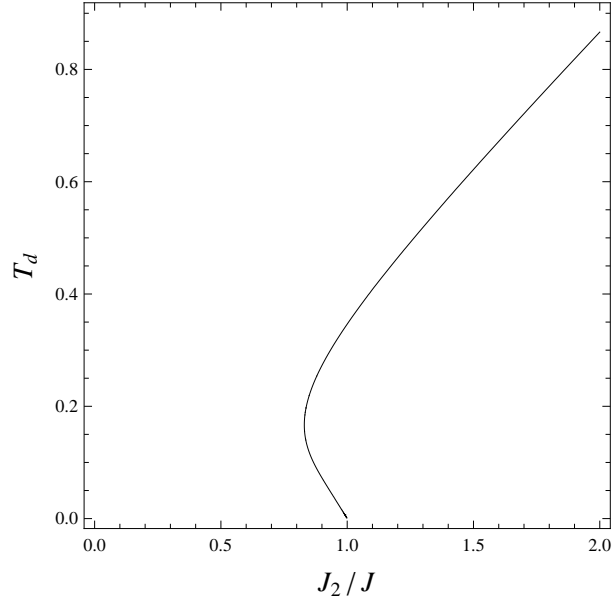}
\caption { \small {Sudden-death temperature $T_d$ corresponding to the vanishing or arising of entanglement at zero magnetic $H$ versus ration parameter $J_2/J$ ($J=0.5$).}} \label{C_H}
\end{center}
\end{figure}

On the other hand, as it can be seen from Fig.~\ref{C_T_0}, there
are two sudden-death temperatures in the case $J-J_2>0$ (corresponding
to arising and vanishing of entanglement) \cite{solomon}. The
dependence of $T_d$ on the ratio parameter $J_2/J$ is shown in
Fig.~\ref{C_H}. In the area $0<J_2/J<1$, there are two sudden-death temperatures (as mentioned above), while for the values
$J_2/J\geq1$, the dependence is a linear one.

Our further investigation concerns the effects of the magnetic
field $H$.

\begin{center}
\begin{figure*}
\begin{tabular}{ c c }
\small(a)  &  \small(b) \\
\begin{figurehere}
\includegraphics[width=6cm]{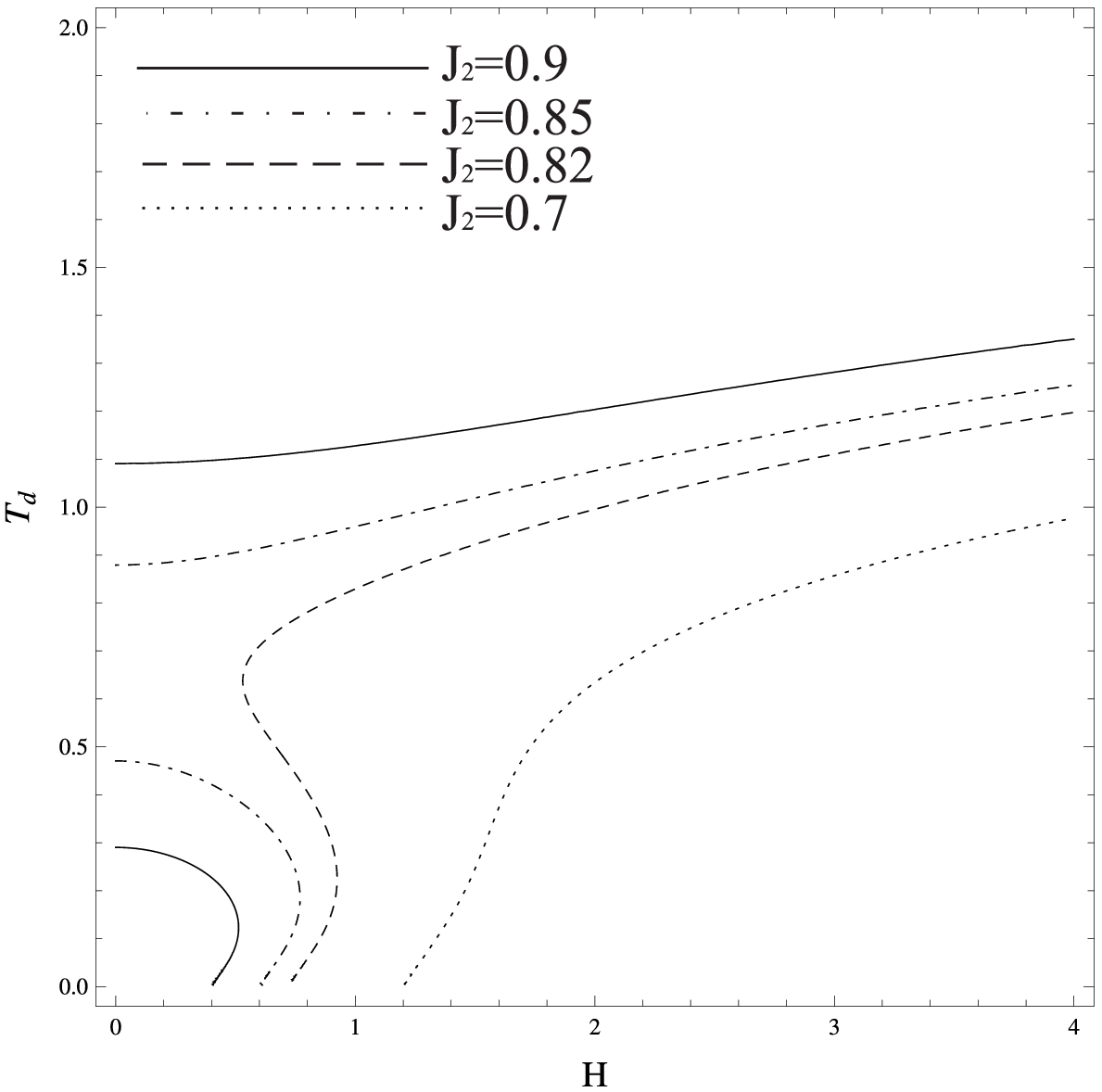}
\end{figurehere} &
\begin{figurehere}
\includegraphics[width=6cm]{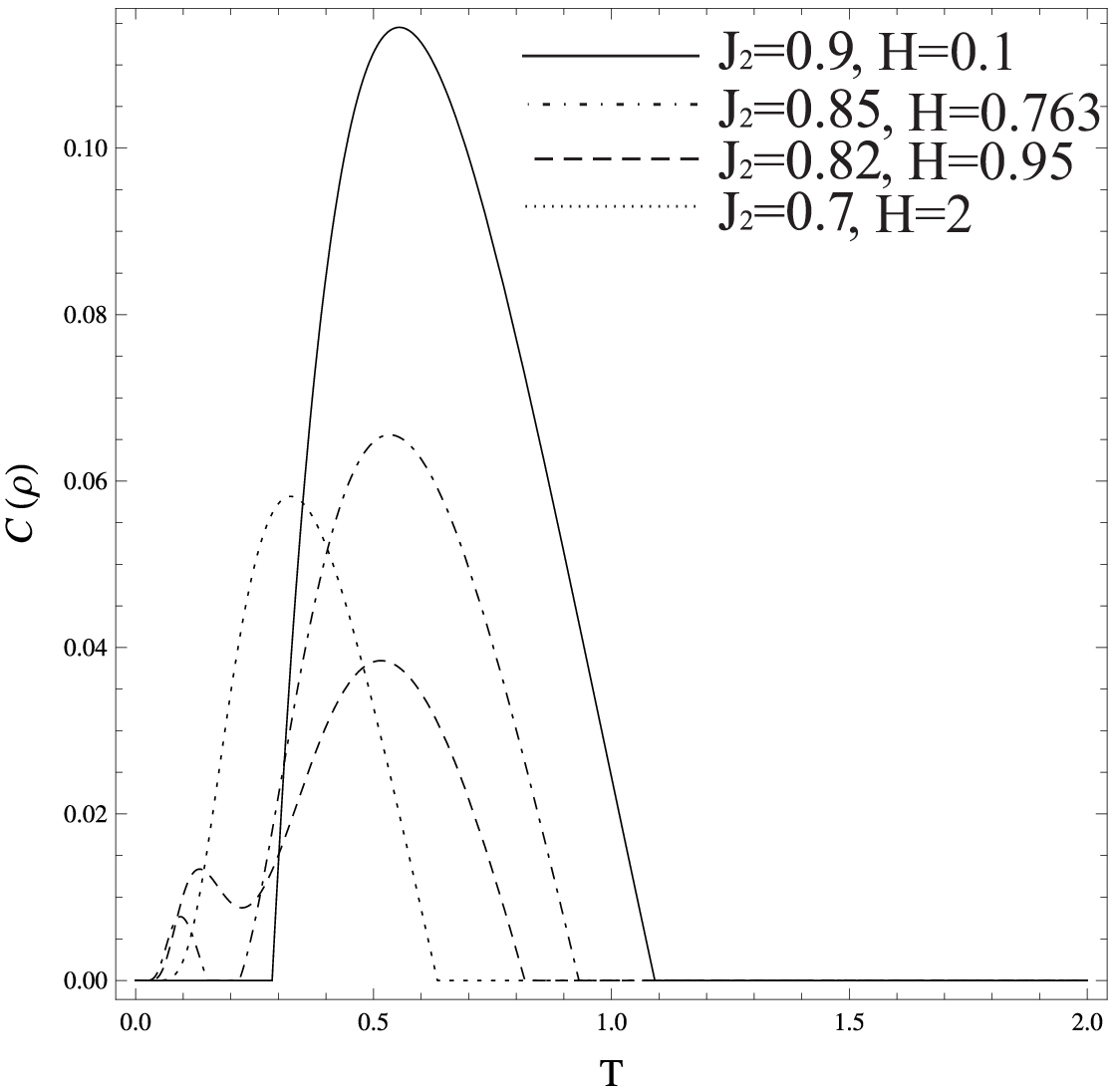}
\end{figurehere}  \\
\end{tabular}
\caption {\small{(a) Sudden-death temperature $T_d$ corresponding to the vanishing or arising of entanglement versus magnetic field $H$ for $J=2$ and different values of $J_2$; (b) Concurrence $C(\rho)$ versus temperature $T$ for $J=2$ and different values of $J_2$ and magnetic field $H$. \label{T_c_h}}}
\end{figure*}
\end{center}

Firstly we will discuss how magnetic field affects the above
introduced sudden-death temperature $T_d$. While increasing $H$, $T_d$
increases too, but it always remains lower than $J_2/\log3$ (in
other words $\mathrm{lim}_{H\rightarrow\infty}T_d=J_2/\log3$).
Another interesting fact is that magnetic field gives a rise to more
than two sudden-death temperatures in the case $J-J_2>0$
[Fig.~\ref{T_c_h} (a)] and on the dependence of $C(\rho)$ from
temperature $T$ one finds two peaks separated by an area of a zero
entanglement [Fig.~\ref{T_c_h} (b)]. With increasing $H$ the smaller
of aforementioned peaks tears apart from $C(\rho)=0$, starts merging
with the bigger one and eventually disappears. An effect of this
kind has not been reported yet, to the best of our knowledge.
Although a similar two-peak behavior of concurrence was found in
dissipative the Lipkin-Meshkov-Glick model versus \textit{magnetic
field} \cite{two}. However, when $T\rightarrow0$, $C(\rho)$ remains
finite and becomes zero only at absolute zero temperature $T=0$
[i.e. there can be not more than three sudden-death temperatures
corresponding to disappearing or arising of thermal entanglement, as
it can be also seen form Fig.~\ref{T_c_h} (a)]. In other words in
the area of low temperatures the behavior of concurrence is smooth,
in contrast with the case when magnetic field is absent.

\begin{figure}
\begin{center}
\includegraphics[width=8cm]{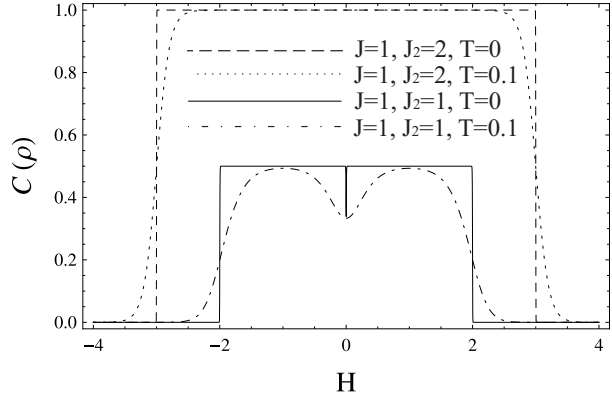}
\caption { \small {Concurrence $C(\rho)$ versus magnetic field $H$ for
different values of temperature, $J_2$ and $J$.}} \label{H}
\end{center}
\end{figure}

Now, we concentrate on the dependence of $C(\rho)$ on magnetic
field. Because of the above introduced ground state structure, the dependency curve of $C(\rho)$ from magnetic field at zero
temperature has a dip at $H=0$ with $C(\rho)=1/3$ for
$J-J_2=0$. There is no dip if $J-J_2<0$  (Fig.~\ref{H}). When
Ising-type interaction is stronger than the Heisenberg one
($J-J_2>0$), one does not find a magnetic entanglement. Furthermore, magnetic
entanglement is of a higher value than that at zero magnetic field
in the case $J-J_2=0$. This happens due to the fact that ground state here is
two-fold degenerated and contains $\psi_{5}$ and
$\psi_{12}$ with the value $C(\rho)=1/2$ for the
corresponding reduced density matrix. $C(\rho)$ becomes zero for the
case $J-J_2\leq0$ at the values of $H$, corresponding to saturation
field, that is when the non-entangled state
$\uparrow\uparrow\uparrow\uparrow$ (in the area $H>0$) or
$\downarrow\downarrow\downarrow\downarrow$ (in the area
$H<0$) becomes the ground state. One can find the described values
of $H$ from the conditions $E_{9}=E_{5}$ and $E_{16}=E_{8}$, giving
$H_s^+=J+J_2$ and $H_s^-=-J-J_2$, respectively. Thermal effects
smoothes the step-like behavior of concurrence in the case when
$J-J_2\geq0$ and induces thermal entanglement when $J-J_2>0$ (see
Fig.~\ref{C_T_0}). The further increase of temperature causes the
quantum correlations eventually dying out for the both cases.

\begin{center}
\begin{figure*}
\begin{tabular}{ c c }
\small(a)  &  \small(b) \\
\begin{figurehere}
\includegraphics[width=6cm]{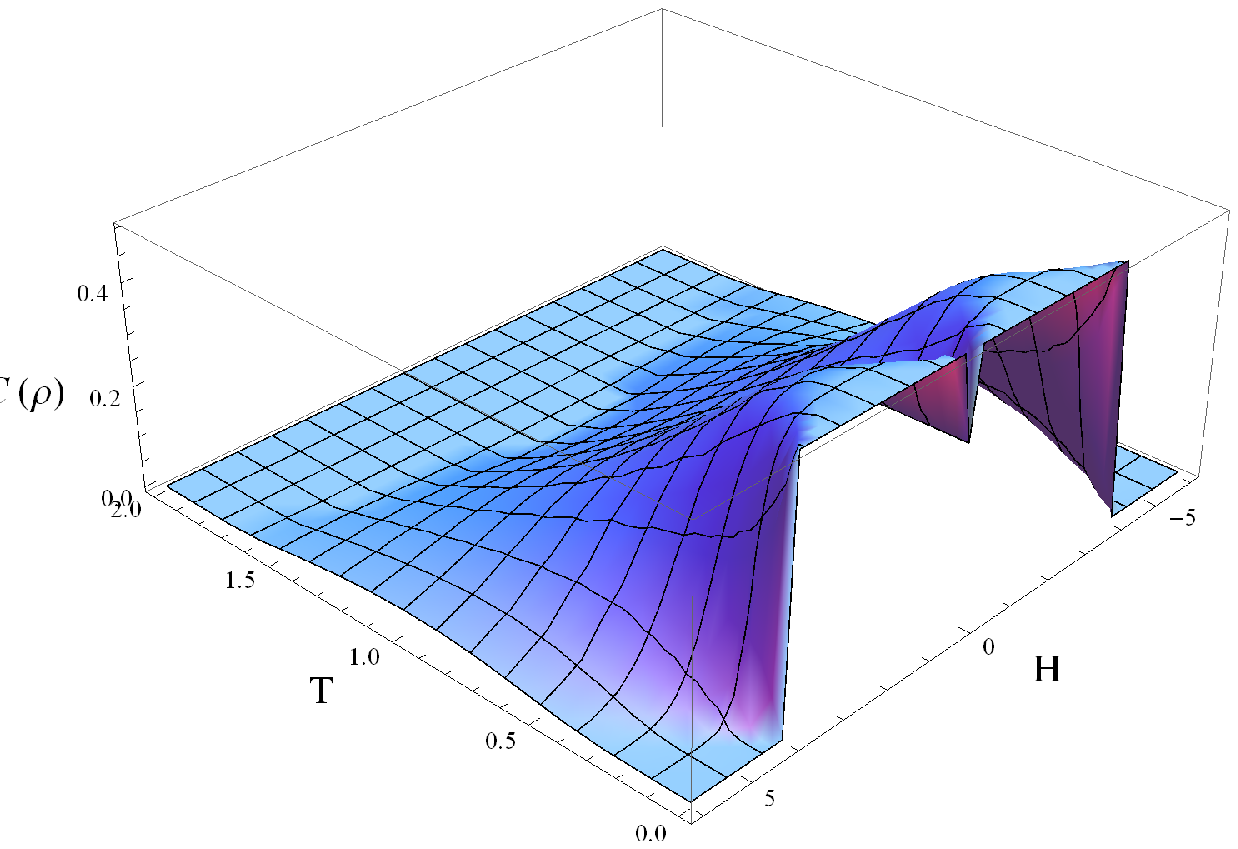}
\end{figurehere} &
\begin{figurehere}
\includegraphics[width=6cm]{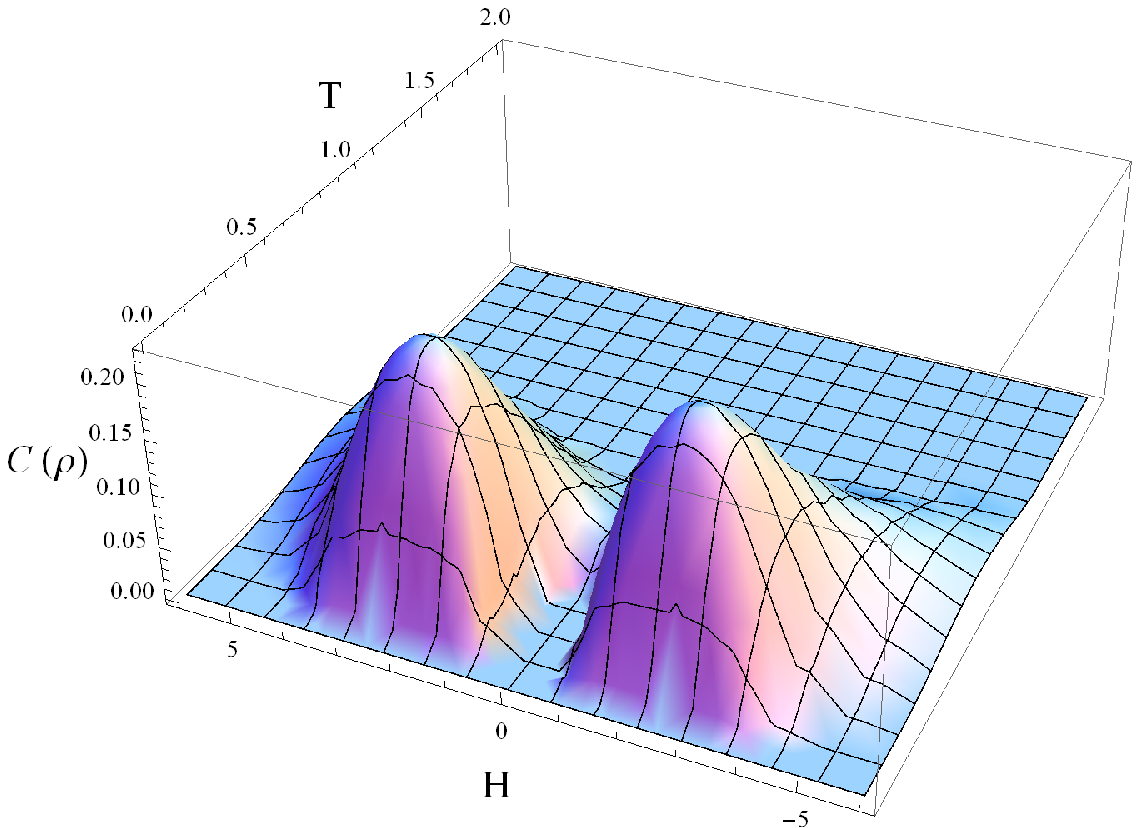}
\end{figurehere}  \\
\end{tabular}
\caption {\small{Concurrence $C(\rho)$ versus magnetic field $H$ and temperature $T$ for (a) $J_2=2$ and $J=2$; (b) $J_2=1.7$ and $J_1=J_3\equiv J=2$. \label{C_H_T}}}
\end{figure*}
\end{center}

Summarizing, in Fig.~\ref{C_H_T} we also plot three-dimensional
dependencies of the concurrence $C(\rho)$ versus temperature $T$ and
magnetic field $H$.

\section{Incorporation of $J_m$ interaction}\label{Jm}

\begin{figure}
\begin{center}
\includegraphics[width=8cm]{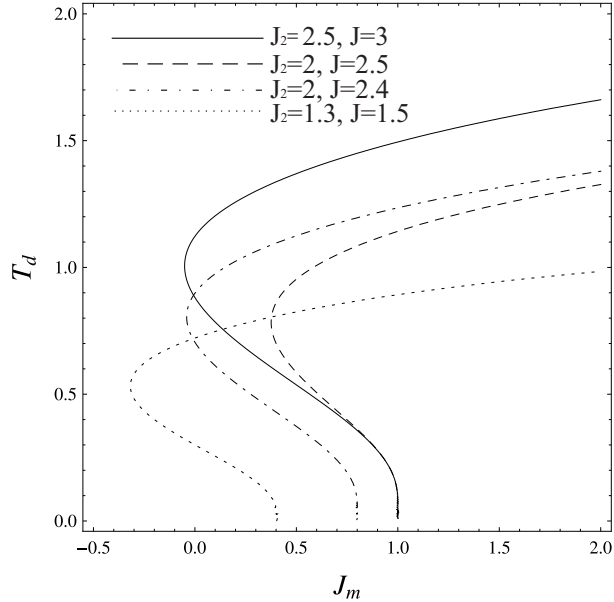}
\caption { \small {Sudden-death temperature $T_d$ corresponding to the vanishing or arising of entanglement versus $J_m$ for $H=0$ and different values of $J_2$ and  $J=2$.}} \label{T_C_m}
\end{center}
\end{figure}

In this section we will study the effects of the next-nearest neighbor
interaction $J_m$ between the Ising spins of the cluster, using the full expression for (10) and (\ref{9}).
We will start with the discussion of the ground state structure for
the case $H=0$ and $J-J_2>0$. It turns out that here one can
distinguish two regimes. First, when $0<J_m<2(J-J_2)$, frustrated
ground state contains two-fold degenerate $\psi_{12}$ and
$\psi_{13}$  and thus the dependency curve of $C(\rho)$ from
temperature starts up at $C(\rho)=0$ . However, the thermal effects
can cause the thermal entanglement for the values of $J_m$, close to
$2(J-J_2)$ (but remaining $J_m<2(J-J_2)$). One finds two sudden-death temperatures on the dependency curve of $C(\rho)$ from $T$
(Fig.~\ref{T_C_m}). This effect can be understood from the following
discussion. The ground state consists of four-fold degenerate states
$\psi_{6}$, $\psi_{7}$, $\psi_{12}$ and
$\psi_{13}$, for the case $J_m=2(J-J_2)$. Although this
mixture contains maximum entangled states $\psi_{6}$ and
$\psi_{7}$, the corresponding density matrix for this ground
state gives $C(\rho)=0$. By increasing temperature, one obtains the
thermal mixing of states which leads to a higher contribution of
entangled states. This contribution, however, becomes less, when the
values of $J_m$ are considerably higher than $2(J-J_2)$. Thus, with
increasing the difference of $J_m$ and $2(J-J_2)$, the local maximum
becomes narrower and eventually disappears.

In the opposite case, when $J_m>2(J-J_2)$, the frustrated ground
state is two-fold degenerate, but with $\psi_{6}$ and
$\psi_{7}e$, hence above mentioned curve of $C(\rho)$ starts
from $C(\rho)=1$. We find only one sudden-death temperature here, which increases with the growth of $J_m$
(Fig.~\ref{T_C_m}). In other words, the qualitative picture remains
the same as for the case $J_m=0$.

$C(\rho)$ is of a maximum value ($C(\rho)=1$) at zero magnetic field
and zero temperature, regardless of the $J_m$ for
a dominant Heisenberg interaction ($J-J_2<0$).

Concluding the discussion of zero magnetic field properties in the
case $J_m\neq0$, we note, that when $J_m<0$
(ferromagnetic coupling), the absolute value of $J_m$ does not
interfere with the ground state properties of the system (it will be
two-fold degenerate $\psi_{12}$ and $\psi_{13}$, if $J>J_2$ or
$\psi_{5}$ and $\psi_{8}$, if $J<J_2$).

Here then, we will discuss the regime $J-J_2>0$ introducing effects
of the magnetic field $H$. We differentiate two subcases. First
one, when $J_m\leq2(J-J_2)$, one does not find magnetic entanglement
in the system, since increasing the absolute value of magnetic field
$H$, we obtain a sequence of separable states (e.g.
$\psi_{12}\rightarrow(\psi_{10}+\psi_{11})\rightarrow\psi_9$
or $\psi_{12}\rightarrow\psi_{9}$ for $H>0$). Here
and further by $(psi_{i}+\psi_j)$ we will mean
two-fold degenerate states.

\begin{figure}
\begin{center}
\includegraphics[width=8cm]{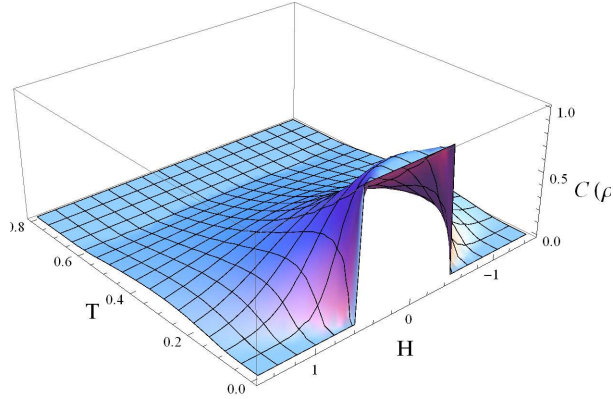}
\caption { \small {Concurrence $C(\rho)$ versus magnetic field $H$
and temperature $T$ for $J_2=1$ and $J=1.5$ and
$J_m=1.5$.}} \label{C_3_1}
\end{center}
\end{figure}

Meanwhile, when $J_m>2(J-J_2)$, the aforementioned sequence
of states starts from $(\psi_6+\psi_7)$ with
maximum value of $C(\rho)=1$, therefore we obtain magnetic
entanglement (Fig.~\ref{C_3_1}). One can introduce here critical
values of magnetic field $H_c^+$ and $H_c^-$, corresponding to
vanishing of magnetic entanglement. In contrary with the case
$J_m=0$, $H_c^{\pm}$ does not coincide with the saturation fields $H_s^\pm$ (see Sec.~\ref{zero}).

\begin{center}
\begin{figure*}
\begin{tabular}{ c c }
\small(a)  &  \small(b) \\
\begin{figurehere}
\includegraphics[width=6cm]{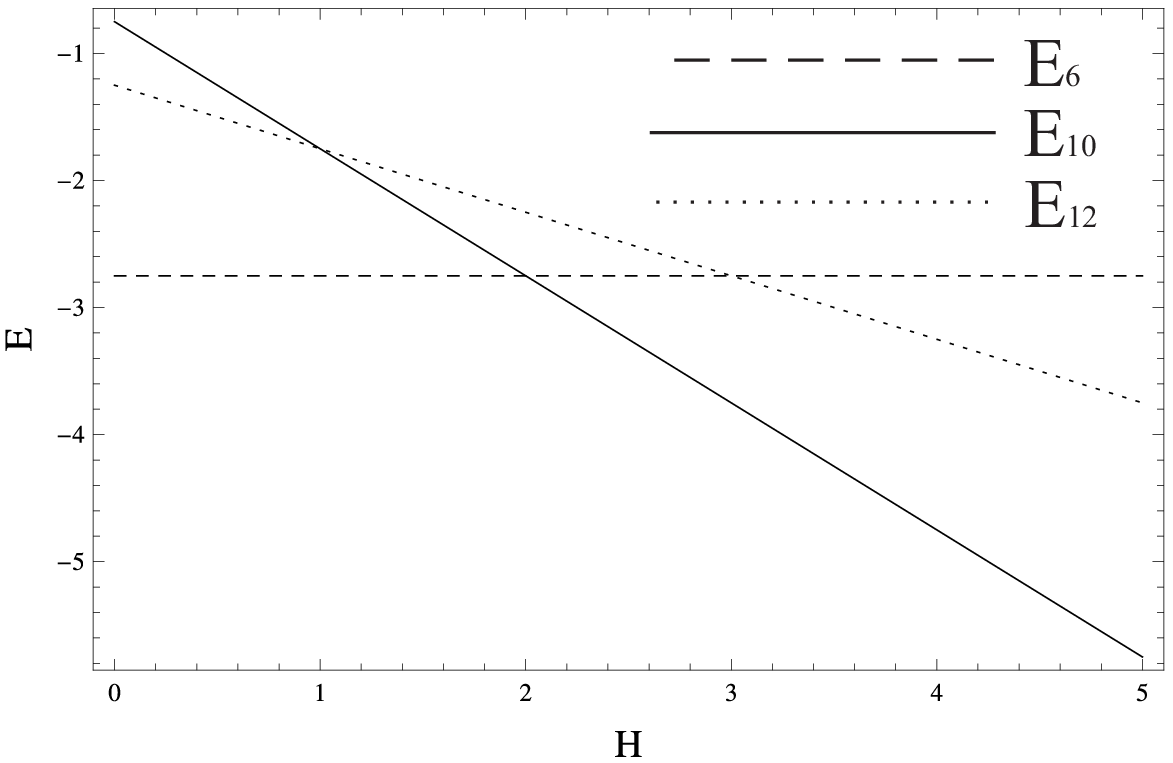}
\end{figurehere} &
\begin{figurehere}
\includegraphics[width=6cm]{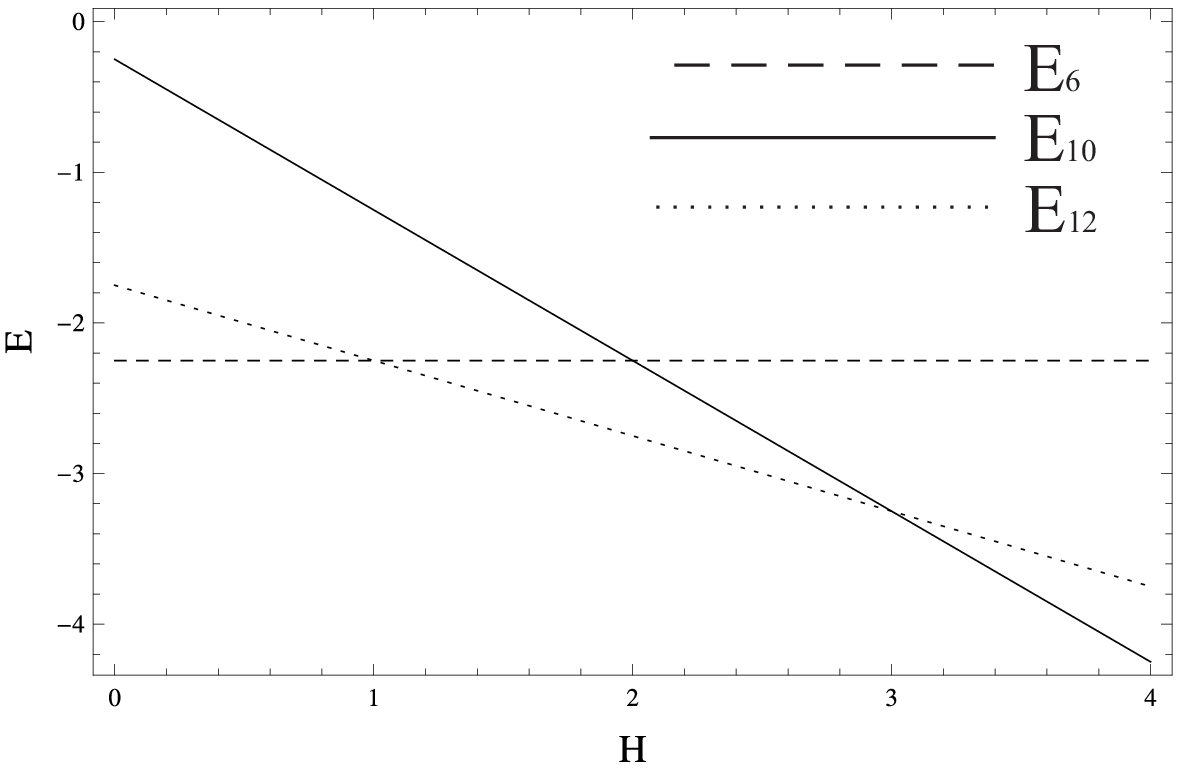}
\end{figurehere}  \\
\end{tabular}
\caption {\small{Eigenvalues $E_6$, $E_{10}$ and $E_{12}$ versus
magnetic field $H$ for (a) $J_m<2J-J_2$; (b) $J_m<2J-J_2$.
\label{enrgy}}}
\end{figure*}
\end{center}

On the one hand, we have the ground state transitions
$(\psi_6+\psi_7)\rightarrow\psi_{12}\rightarrow(\psi_{10}+
\psi_{11})\rightarrow\psi_{9}$
(for $H>0$) if the value of magnetic field, corresponding to the
intersection of energies $E_{6}$ and $E_{12}$ is lower than that of
$E_{10}$ and $E_{12}$ [Fig.~\ref{enrgy} (a)]. This condition gives:
$J_m<2J-J_2$. Thus the corresponding critical values of magnetic
field can be found from $E_6=E_{12}$ with $H_c^+=2J_2 - 2J + J_m$
(obviously, $H_c^-=-H_c^+$, from equation $E_{6}=E_{13}$). On the
other hand, when $J_m>2J-J_2$, we have the ground state transitions
$(\psi_{6}+\psi_{7})\rightarrow(\psi_{10}+\psi_{11})\rightarrow\psi_{9}$
[Fig.~\ref{enrgy} (b)]. Corresponding $H_c^+=J_2$, found from
$E_6=E_{10}$ ($H_c^-=-J_2$, from $E_{6}=E_{14}$).  The ground state
transition
$(\psi_{6}+\psi_{7})\rightarrow\psi_{9}$ can
not occur, since the corresponding condition is inconsistent
with $J_m>2J-J_2$.

Following the same technique as in previous paragraph (we will not
stop on detailed phase structure), we obtain the following regimes
for $J-J_2<0$: $H_c^+=H_s^+=J+J_2$ ($H_c^-=-H_c^+$) in the case
$J_m<J_2-J$ and $H_c^+=2J_2-J_m$ if $J_2>J_m>J_2-J$, and finally,
when $J_m>J_2$ one finds $H_c^+=J_2$.

\begin{figure}
\begin{center}
\includegraphics[width=8cm]{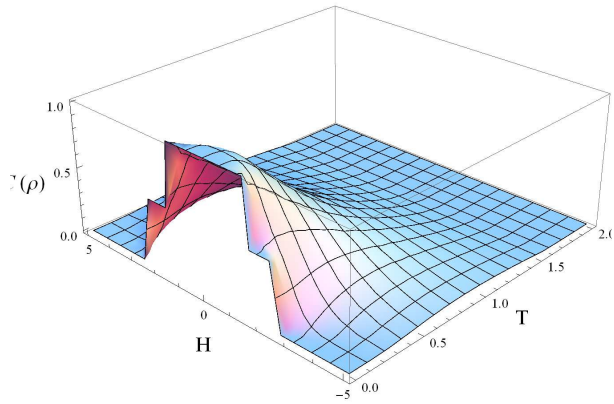}
\caption { \small {Concurrence $C(\rho)$ versus magnetic field $H$
and temperature $T$ for $J_2=2$ and $J=2$ and
$J_m=1.5$.}} \label{C_3_2}
\end{center}
\end{figure}

The special (boundary) case $J-J_2=0$ is also of interest, since one
can observe here magnetic entanglement of different values
($C(\rho)=1$ and $C(\rho)=1/2$) (Fig.~\ref{C_3_2}), whereas in the
case $J_m=0$, these two regimes cannot coexist for a fixed values of
$J$ and $J_2$. This situation arises only for $0<J_m<J$, when one finds the
 sequence of states $(\psi_{6}+\psi_{7})\rightarrow(\psi_{5}+\psi_{12})\rightarrow$(factorable state) (for $H>0$).
In other words, at the values of magnetic field $H=\pm J_m$ (found
from conditions $E_{6}=E_{12}$ for $H>0$ and $E_{6}=E_{8}$ for $H<0$) the
states with different
 values of magnetic entanglement coexist.

As for the sudden-death temperature corresponding to the disappearing or
arising of entanglement at non-zero magnetic field, one finds a
similar behavior as in the case $J_m=0$, i.e. here again we find up
to three sudden-death temperatures (as in Fig.~\ref{T_c_h}), with a two
peak behavior on the dependency of $C(\rho)$ on temperature.


\section{Conclusion}\label{concl}

In this paper we have studied the thermal entanglement of a spin-1/2
Ising-Heisenberg model on a symmetrical diamond chain, which has
been proposed to understand a frustrated magnetism of the series of
compounds, like A$_3$Cu$_3$(PO$_4$)$_4$ with A=Ca, Sr,
Bi$_4$Cu$_3$V$_2$O$_{14}$, Cu$_3$(TeO$_3$)$_2$Br$_2$  and
Cu$_3$(CO$_3$)$_2$(OH)$_2$. We have studied the phase structure and
entanglement properties of the system in a wide range of Ising-type
interaction constants $J_1=J_3\equiv J$, $J_m$ and Heisenberg-type
$J_2$, considering that diamond chain structure describes a broad
class of materials (within different values of exchange interaction
parameters) and that the exact value of coupling constants for
azurite (Cu$_3$(CO$_3$)$_2$(OH)$_2$) is still under scrutinizing
question. Taking into account the classical and hence separable
character of Ising-type interactions which are coupling adjacent
Heisenberg dimers, we have calculated the entanglement of each of
these dimers separately. We have used the concurrence for
quantifying the amount of entanglement between two Heisenberg-type
spins, by tracing out Ising-type ones from the density matrix of the
diamond-shaped cluster (the only entangled pair here is the
Heisenberg dimer). The incorporation of next-nearest neighbor
interaction $J_m$ has also been investigated (generalized diamond
chain) and the effects of external magnetic field have been invoked.

We have revealed a number of regimes with distinct ground state
structure and qualitatively different thermodynamic behavior,
depending on the relations between $J, J_2$ and $J_m$ and values of magnetic
field $H$. We found that in general for a dominant Heisenberg-type
interaction ($J_2>J$) the system's ground state is maximally
entangled, but increasing the temperature, pure quantum correlations
eventually disappears. On the other hand, for a dominant Ising-type
interaction ($J>J_2$) the ground-state is non-entangled, whether the
temperature gives rise to thermal entanglement. In the latter case
one does not find magnetic entanglement at the absolute zero
temperature (the system behaves as a classical one). However,
magnetic field can lead to another, yet not described effect of
two-peak behavior of concurrence $C(\rho)$ versus
\textit{temperature} with three sudden-death temperatures (one of them
corresponding to reappearance of concurrence and the other two to
its disappearing). These two peaks are separated by an area of a
zero entanglement, which becomes narrower, with the growth of the magnetic
field and aforementioned peaks merge into each other. Another novel
effect was indicated for a boundary case $J=J_2$ when $0<J_m<J$.
Specifically, two states with different values of magnetic
entanglement coexist for the value of magnetic field $H=\pm J_m$. One finds a step like behavior of concurrence versus magnetic field
$H$, with plateaus at the value 1/2 and 1. In other words, the
presence of competing interactions in the system and geometrical
structure of the chain, each leading to a frustration, makes the
phase structure of the system richer and gives rise to an
interesting physical behavior. Finally, the adopted model guaranties
an experimental realization for suitable theoretical treatment. Our results will be useful for further experimental detection of
entanglement in the diamond chain structured macroscopic samples by
means of entanglement witnesses (e.g. built from measurements of
magnetic susceptibility \cite{witness}).

\section{Acknowledgments}

The work was supported by the French-Armenian No CNRS IE-017, Brazilian No CEX-BPV-00028-11, ECSP-09-08-SASP NFSAT, PS-2497 ANSEF grants, Marie Curie IRSES SPIDER, Project - PIRSES-GA-2011-295302 (N.A.). L. C. wishes to acknowledge the support of the Regional Council of Burgundy (Conseil R\'{e}gional de Bourgogne), and the International Associated Laboratory IRMAS. O. R. thanks the CNPq for the partial financial support.

\end{document}